# An NFV and Microservice Based Architecture for on-the-fly Component Provisioning in Content Delivery Networks

Narjes Tahghigh Jahromi[†1], Roch H. Glitho[†2], Adel Larabi[‡1], Richard Brunner[‡2]
[†] CIISE, Concordia University, Montreal, QC, Canada
[‡] Ericsson, Montreal, QC, Canada
[†1] n_tahghi@encs.concordia.ca, [†2] glitho@ciise.concordia.ca, [‡1] adel.larabi@ericsson.com, [‡2] richard.brunner@ericsson.com

*Abstract*— Content Delivery Networks (CDNs) deliver content (e.g. Web pages, videos) to geographically distributed end-users over the Internet. Some contents do sometimes attract the attention of a large group of end-users. This often leads to flash crowds which can cause major issues such as outage in the CDN. Microservice architectural style aims at decomposing monolithic systems into smaller components which can be independently deployed, upgraded and disposed. Network Function Virtualization (NFV) is an emerging technology that aims to reduce costs and bring agility by decoupling network functions from the underlying hardware. This paper leverages the NFV and microservice architectural style to propose an architecture for on-the-fly CDN component provisioning to tackle issues such as flash crowds. In the proposed architecture, CDN components are designed as sets of microservices which interact via RESTFul Web services and are provisioned as Virtual Network Functions (VNFs), which are deployed and orchestrated on-the-fly. We have built a prototype in which a CDN surrogate server, designed as a set of microservices, is deployed on-the-fly. The prototype is deployed on SAVI, a Canadian distributed test bed for future Internet applications. The performance is also evaluated.

## I. INTRODUCTION

Over the past decades, the importance of video services has increased considerably. A Cisco Virtual Networking Index forecast [1] predicts that the IP video traffic will take up to 82 percent of all the Internet consumer traffic by 2020. The same report foresees that Content Delivery Networks (CDNs) will carry about two-thirds of all the Internet video traffic by 2020. CDNs have gained immense popularity for their ability in the efficient delivery of content services to a large number of geographically distributed end-users. CDNs are designed as an overlay network of surrogate servers, a.k.a replica or edge servers and a controller [2]. The original contents are replicated from content providers' origin servers to the geographically distributed surrogate servers closer to end-users. This means a significant reduction in content delivery latency and reduction of the load on the core network.

In some events, some content might attract the attention of a large number of end-users and this might lead to flash crowds [3], which are the sudden, unpredicted surges in requests toward particular contents. Flash crowds eventually cause an outage in CDNs, i.e. service unavailability, site slowdowns, and a decrease in Quality of Service (QoS). The terrorist attacks of September 11, 2001, is an example [4].

In case of flash crowds, traditional CDNs are challenged to extend their coverage by either provisioning additional surrogate servers or by increasing the capacity of existing surrogate servers. This paper focuses on the first alternative. Traditionally, CDN surrogate servers are provisioned at fixed network locations and on a proprietary/dedicated hardware. The traditional mode of provisioning surrogate servers takes a significant amount of time and its shortcomings are widely known [5][6].

This paper uses microservice architectural style [7] integrated with NFV technology [8] to enable the on-the-fly provisioning of CDN components including surrogate servers. Microservice architectural style [7] is an approach in which an application or functionality is composed of several smaller and loosely coupled components called microservices. NFV [8] is a novel way to virtualize network services and aims at decoupling the network functions from the underlying hardware. This paper proposes and validates an architecture in which CDN components (e.g. surrogate servers) are designed as microservices which are deployed and orchestrated using NFV principles. In addition, a pre-deployment and a post-deployment phases are considered. During the pre-deployment phase, the optimal Point of Deployments (PoD) are selected according to their characteristics and geographic locations. During the post-deployment phase, the newly deployed CDN component will be filled with contents and will be integrated with the existing CDNs.

Section II introduces a use case, requirements and discusses the related work. Section III is devoted to the proposed architecture. Section IV presents the validation prototype and the evaluation experimentations. Section V concludes the paper and discusses the lessons learned.

## II. USE CASE, REQUIREMENTS & RELATED WORKS

### A. Use Case

Let us consider a business model with a content provider, a CDN provider and eventually an Internet Service Providers (ISP). The content provider, e.g. Canadian Broadcast Corporation (CBC) news, provides contents to end-users. CDN provider operates and manages a CDN and provides content delivery services to content providers. ISP provides Internet connectivity services to end-users and additionally, we assume that it provides PoDs to CDN providers. PoDs are assumed to be located at ISP premises and offer the required

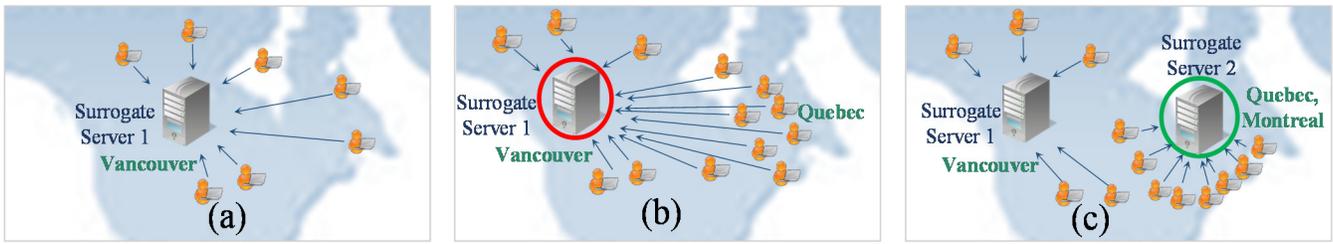

Fig.1. An illustration of flash crowds use case. a) Normal situation; end-users are redirected to surrogate server 1 in Vancouver
b) Flash crowds; a content gains sudden popularity c) Automatic provisioning of a new surrogate server 2 in Montreal and redirection of end-user requests.

infrastructure and resources for CDN component provisioning.

Let us assume the CBC contents are replicated on a surrogate server in Vancouver and as shown in Figure 1.a, in normal cases, the end users in Canada area are redirected to this surrogate server. Let us now assume that an unpredicted event (e.g. Quebec floods in 2017) happens in which many contents are distributed and some of them gain popularity in short period of time. The CDN provider notices a surge of traffic from a given area (e.g. Quebec area) which is covered by an ISP. As shown in Figure 1.b, the massive increase in requests for the contents might lead to flash crowds. The CDN provider contacts the ISP in an automated manner to request the deployment of a new CDN surrogate server in Quebec area. ISP will provide high-level information of their available PoDs and after the optimal PoD is selected, it will allow the deployment of a CDN surrogate servers in Montreal. Afterward, the newly deployed surrogate server will be filled with the popular contents related to the event and will be integrated with the existing network of CDN provider in an automatic way. After the successful process of surrogate server provisioning, as illustrated in Figure 1.c, the end-users in Quebec area will be automatically redirected to the new surrogate server. CDN provider might dispose their content delivery services after the number of requests decreases.

*B. Requirements*

The first requirement is the ability to cover the whole provisioning process. It should be noted that the provisioning process includes pre-deployment, actual deployment, and post-deployment processes. An example of the pre-deployment process is selecting the appropriate PoD for surrogate server provisioning according to PoD specifications and content provider requirements. Deployment process includes the instantiation and activation of CDN component e.g. surrogate server functionality on selected PoDs. The post-deployment process includes but not limited to integration of newly deployed surrogate server into the existing CDNs.

The second requirement is that the provisioning process should be fully automated. This is realized by automating all the provisioning phases including pre-, post-, and actual deployments.

The third requirement is that CDN architecture should allow provisioning the CDN surrogate servers on heterogeneous underlying resources.

The fourth requirement is that the architecture should be based on well-adopted technologies and widely deployed standards. Examples are cloud computing, NFV, etc. This requirement eventually will ensure that the architecture enables interoperability with external domains and entities.

*C. Related Works*

To the best of our knowledge, no paper has attempted so far to address the problem of on-the-fly provisioning of CDN components using NFV and microservice architectural style. Some papers focus on deployment of application components in general, while others focus on various methods for CDN component deployments.

*1) Component Deployment in General*

The European Telecommunications Standards Institute (ETSI) has defined a reference architectural framework for NFV [9] which enables deployment of application components in general. As shown in Fig. 2, the NFV framework includes three main modules: VNFs, NFV Infrastructure (NFVI), and NFV Management and Orchestration (MANO). A VNF is the software implementation of a given functionality. NFVI provides the virtualized and physical resources as services. NFVI is managed by NFV MANO, providing the required environments for VNFs to be deployed and executed. NFV MANO is also responsible for VNF lifecycle management and service provisioning including deployment, execution, and management. This work partially satisfies our first and second requirements, due to the fact that it solely focuses on automatic deployment and orchestration of application components and overlooks the post-deployment processes. On the other hand, this work satisfies our third and fourth requirements, since it uses NFV as a well-adopted technology and enables deployments on heterogeneous devices.

Active networking [10] is another technology which allows the service providers, deploy customized components in form of pieces of software on network elements such as routers and switches. This leads to making the network elements active, in terms of performing computations within the network. Examples of deployed components include firewalls, gateways, video mixers, etc. This technology cannot fully meet our first and second requirements, mainly because, although it enables the automatic deployment of components, it doesn't consider the problem of pre- and post-provisioning of components in various contexts. When it comes to the third requirement, it enables deployment on heterogeneous devices as long as the deployment platform is provided. This technology doesn't meet our fourth requirement since active networking is not a widely-adopted and implemented paradigm.

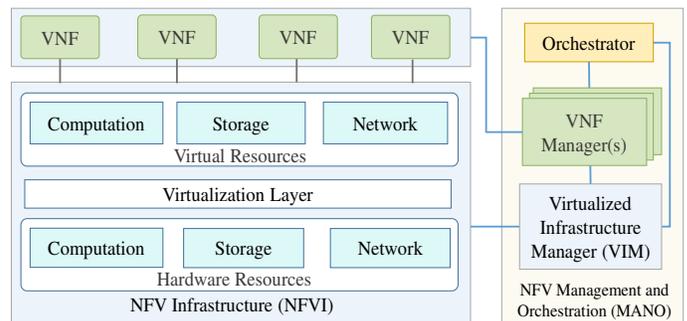

Fig.2. A high-level view of the NFV architecture

*2) CDN Component Deployment*

In CDN context, ETSI has introduced an NFV architecture use case for CDNs [6]. It proposes the virtualizing of the CDN surrogate servers in form of VNFs and motivates the need for CDN providers to dynamically deploy and manage surrogates in a telecommunication network operator domain. However, it overlooks the problem of pre- and post-deployment of surrogate servers and doesn't fully satisfy our first and second requirements. It should be noted that we use this work as the baseline and build our proposed architecture by offering extensions to NFV MANO.

Ref [11] proposes an architecture for CDN as a service (CDNaaS), which allows content providers to order and deploy the CDN surrogate servers in ISPs. ISPs leverage CDNaaS to receive the content provider requests, orchestrate the resources, and deploy the surrogate server functionality on available infrastructure. This architecture partially satisfies our second requirement as it focuses on automation of pre-deployment and deployment phases, but it doesn't consider any method for post-deployment processes.

Another related work is ActiveCDN [5] that allows content providers to dynamically scale content delivery services by utilizing virtualization on storage resources of network elements. This architecture partially satisfies our first and second requirements as it solely focuses on deployment phase in which a network virtualization framework, called NetServ [12] is used to allow network service deployments on-the-fly. ActiveCDN is powered by active networking technology [13], meaning that network functionalities can dynamically be created and perform customized computations on the packet content. However, since Active networking is not a widely-deployed technology, this work does not meet our fourth requirement.

Netflix [14] is a commercial related work which is basically a content provider that relies on its own CDN infrastructure. Netflix relies on NGINX web server which is designed based on microservice architectural style. Netflix surrogate servers, called Open Connect Appliance (OCA), are hardly coupled with dedicated proprietary hardware. Netflix partners with ISPs, and afterward, ships and deploys the OCA devices to their premises. Netflix doesn't meet any of our requirements as the OCA relies on dedicated hardware and proprietary technologies, and the deployment and configuration of OCAs are manual and require labor-intensive tasks.

## III. ARCHITECTURE FOR ON-THE-FLY PROVISIONING OF CDN SURROGATE SERVERS

In this section, the architectural principles are first introduced. Second, the high-level architecture of the designed system, including its related architectural planes, components, and interfaces, are discussed. Third, an illustrative sequence diagram is discussed.

### A. Architectural Principles

The first architectural principle is the design of each CDN component, e.g. surrogate server, as a set of microservices and each microservice focuses on completing an independent task. The second architectural principle is that the REpresentational State Transfer (REST) [15] architectural style is used for designing the interactions between the CDN component microservices. REST is selected because it is standard-based, lightweight and flexible for data representation, and allows us to describe APIs in a generic and abstract way. The third architectural principle is the packaging of CDN component microservices as VNFs, and the use of NFV technology for their automatic deployments.

### B. High-Level Architecture

A high-level view of the designed architecture is depicted in Fig. 3. It should be noted that the business model we propose in this article, includes a new actor, CDN component provider, in addition to traditional CDN business model actors. The CDN component provider provides CDN component functionalities to CDN providers. Each functionality is packaged in form of VNF and is deployed by CDN component provider on the selected PoDs. The PoDs might be owned and managed by either ISPs or CDN providers. However, they offer well-defined APIs to the CDN component providers to deploy the CDN components there. The proposed architecture is layered over a set of planes. These planes and the contained components are discussed below. We also discuss the interfaces between the components in this section.

*1) Architectural planes*

The proposed architecture consists of control and data planes. The control plane handles the signaling between architecture components for ordering a new deployment, exchanging information e.g. CDN components and PoD specifications, requesting deployment and orchestration of microservices, and integrating the new surrogate server in existing CDNs. The data plane provides required resources/environment which allows the actual deployment, orchestration, content exchange and content delivery tasks.

*2) Architectural components*

In what follows, we mainly focus on the components of the control plane since they are the innovative components in this architecture. The architectural components are categorized domain-wise, into CDN provider domain and CDN component provider domain categories.

*a) Components in CDN Provider Domain*

*CDN controller* is in charge of integration of newly deployed nodes into the existing CDNs, in addition to its traditional functionality which is redirecting the end-user requests to the appropriate surrogate server. *CDN deployment manager* is responsible for choosing the PoD, and initiating the request for CDN components provisioning to CDN component provider domain. It will be triggered by event handlers which predict flash crowds or by coverage extension requests from content provider domain.

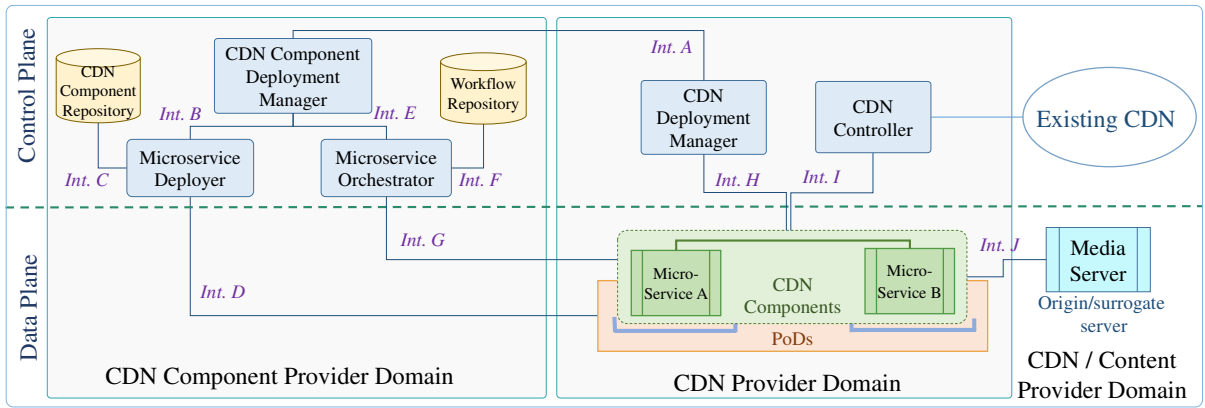

Fig.3. High-level Architecture

*b) Components in CDN component Provider Domain*

*CDN component deployment Manager* is responsible for analyzing the features of requested CDN component and decomposing the request into a set of microservices and the way they should be orchestrated. *CDN component repository* holds the software implementation of various components of a surrogate server packaged in form of VNFs. *Microservice deployer* has the role of deploying the required CDN component microservices on the pre-selected PoDs. In NFV MANO architecture, this functionality is done by the VNF manager. *Workflow repository* holds orchestration plans in form of pre-defined workflows, designed for orchestration of CDN component microservices. *Microservice orchestrator* is responsible for the orchestration of newly deployed microservices according to pre-defined orchestration plans, in a way that finally they interact and function as a full-fledged surrogate server.

It should be noted that the NFV orchestrator in MANO architecture (Fig. 2) performs both resource and network service orchestrations. In our proposed architecture, resource orchestration is part of the *CDN deployment manager* responsibilities, while the *microservice orchestrator* performs the network service orchestration.

*3) Interfaces*

In what follows, the main interfaces are described. Some of the interfaces are inter-domain interfaces which enable the interactions between business domains, while others are intra-domain ones. It should be noted that all the control interfaces are designed based on the RESTful paradigm.

*a) Inter-domain Interfaces*

Int. A is the main signaling interface between the CDN provider domain and CDN component provider domain for ordering component provisioning. This interface is used by *CDN deployment manager* to, firstly, fetch the CDN component catalogue from the *CDN component deployment manager*. The catalogue is offered by CDN component provider domain and includes the list of various available CDN components, along with their specifications. Secondly, the *CDN deployment manager* browses and selects the available CDN components that best match to their needs, and order their provisioning on pre-selected PoDs. The order contains the selected CDN component identifier (ID) and PoDs access information. Table I shows the three REST API offered through this interface, including resources, supported operations, required request parameters and the response content of each operation.

Int. D is a REST-based data path interface (between the microservice deployer and the PoD deployment agent) and is used for deploying the microservices over PoDs. Int. G is another REST-based data path interface (between the microservice orchestrator and newly deployed microservices) and is used for orchestrating the microservices, i.e. sending the access information of other microservices.

*b) Intra-domain Interfaces*

*i) Interfaces in CDN Provider Domain:*

Int. H (between the CDN deployment manager and newly deployed surrogate server) is for sending the access information of the CDN controller that the newly deployed node should be registered in. Int. I (between the newly deployed surrogate server and *CDN controller*) is for registering the newly deployed surrogate server into the CDN controller to eventually be integrated into the existing CDN. This interface is used by the CDN controller to send the list of contents and their locations (an origin server or another surrogate server) to the newly deployed surrogate server to download the content. Int. J (between the newly deployed surrogate server and the *media server*) is for pulling the content from existing surrogate servers or origin server.

TABLE I. EXAMPLES OF THE API OPERATIONS EXPOSED BY THE CDN COMPONENT PROVIDER DOMAIN

| REST Resource | Operation | HTTP Action and Resource URI | Request Body Parameters | Most Important Info in the Response |
|---|---|---|---|---|
| CDN component catalogue | Fetch CDN component catalogue | GET: /CDNComponentCatalogue | None | List of CDN component types and their features |
| CDN components | Provision a CDN component on a PoD | POST: /CDNComponent/{CDNComponentTypeID} | PoD access information | CDN component ID |
| | Dispose a CDN component | DELETE: /CDNComponent/{CDNComponentID} | CDN component ID | Success/failure |

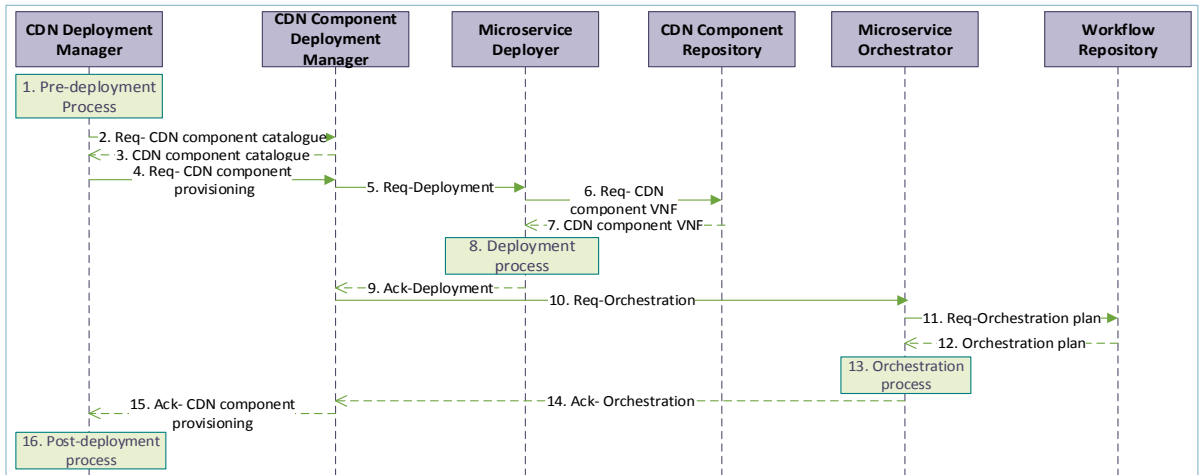

Fig. 4. Sequence diagram of on-the-fly CDN component provisioning

*ii) Interfaces in CDN Component Provider Domain:*

Int. B (between the CDN component deployment manager and microservice deployer) is for requesting the deployment of a set of CDN component microservices. Int. C (between the microservice deployer and CDN component repository) is for fetching the required CDN component microservices. Int. E (between the CDN component deployment manager and microservice orchestrator) is for requesting the orchestration of newly deployed CDN component microservices. Int. F (between the microservice orchestrator and workflow repository) is for fetching the required orchestration plans.

*4) Illustrative sequence diagram*

The envisioned scenario includes a CDN provider who notices a surge of traffic from a given area and decides to extend their CDN coverage by provisioning new surrogate servers in some locations. Figure 4 depicts a sequence diagram showing the interaction of the architectural components to enable this scenario. Firstly, *CDN deployment manager* performs some pre-deployment processes to select the optimal PoD (Fig.4, action 1). The PoDs might be owned and managed by CDN provider or by third parties such as ISPs. The *CDN deployment manager* sends a request to *CDN component deployment manager*, requesting the CDN components catalogue (action 2). The *CDN deployment manager* receives the catalogue (action 3), selects the best CDN component that matches to their need, according to the required features and QoS and requests the CDN component provisioning (action 4). An example of selected CDN component might be a surrogate server with Adaptive Bit Rate (ABR) streaming feature. In addition to the selected CDN component, the CDN component provisioning request also indicates the access information of the selected PoD. The *CDN component deployment manager* decomposes the CDN component provisioning request into a set of microservices and the way they should be orchestrated. Then it sends a deployment request to *microservice deployer* (action 5). The request includes the required CDN microservices to be deployed. Microservice examples for a surrogate server with ABR streaming feature includes cache node microservice and HTTP-DASH streaming server. The *microservice deployer* downloads the microservice VNFs from *CDN component repository* (action 6, 7), and deploys the VNFs on the pre-selected PoDs (action 8). The *CDN deployment manager* receives the acknowledgment of deployment including the newly deployed microservices access information (action 9) and initiates a request to the *microservice orchestrator* to orchestrate the newly deployed microservices (action 10). The *microservice orchestrator* fetches the appropriate orchestration plan from *workflow repository* (action 11, 12), and orchestrate the newly deployed CDN components in order to be able to interact and act as an ABR surrogate server (action 13). Upon receiving the orchestration acknowledgment (action 14), the *CDN component deployment manager* sends the acknowledgment of surrogate server provisioning to the *CDN deployment manager* (action 15). Eventually, the *CDN deployment manager* starts the post-deployment process (action 16).

Figure 5 depicts a sequence diagram of the Post-provisioning and content delivery phases. The *CDN deployment manager* sends the access information of the designated *CDN controller* to the newly deployed surrogate server (Fig. 5, action 1), in order to get registered and be integrated with the existing CDN. The newly deployed surrogate server will send a registration request to the CDN controller (action 2). The *CDN controller* will decide about the contents which should be located on the new surrogate by running the content placement algorithms (action 3). Afterward, it will send back the access information of the optimal media server (it can be an origin or surrogate server) along with the list of contents (action 4). The newly deployed surrogate server will pull the content from the designated media server (action 5, 6) and will notify the CDN controller to start the request redirection process (action 7). Afterward, when an end-user requests a content, it firstly hits the CDN controller (action 8). The CDN controller runs the request redirection algorithms (action 9) to choose the optimal surrogate server and redirects the end user to the newly deployed surrogate server (action 10, 11) to download the content (action 12).

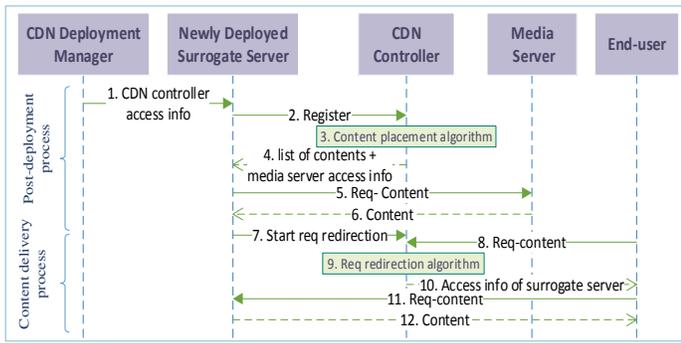

Fig. 5. Sequence diagram of post-provisioning

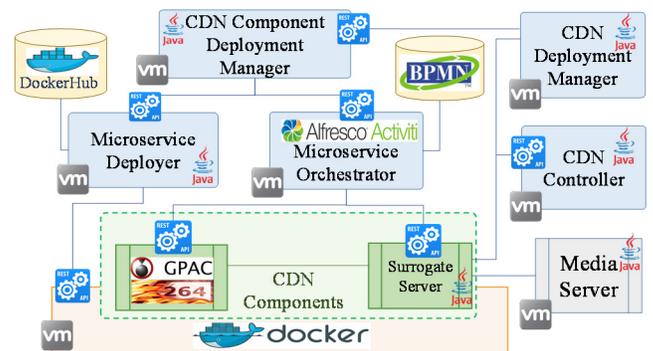

Fig.6. Implementation Architecture

IV. IMPLEMENTATION & EXPERIMENTATIONS

This section details the implementation and the experimentations we performed in order to validate and evaluate our findings. We first present the developed prototype architecture. Then, we present and comment on the experiments we performed to evaluate the performance when provisioning CDN components using our prototype.

*A. Validation Scenario*

The envisioned validation scenario includes a CDN provider who decides to provision a new surrogate server with ABR streaming functionality in Vancouver area. The sequence diagram depicted in Figure 4 is partially implemented for experimentations. It should be noted that the implementation of the full provisioning process requires execution of algorithms for surrogate server placement, content placement, and request redirection which are not the focus of this work. We also assume the CDN catalogue is fetched in advance and the appropriate CDN component is already chosen. Therefore the implemented scenario includes the actions starting from the time the CDN component is requested by CDN provider domain (Fig. 4, action 4), to the time the newly deployed CDN component is registered in the CDN controller successfully (Fig. 5, action 2).

*B. Implementation architecture*

The validation prototype was implemented according to the architecture depicted in Figure 6. All the components are deployed on Virtual Machines (VMs) provisioned using OpenStack IaaS Manager. The latter is provided by Smart Applications on Virtual Infrastructure (SAVI) testbed[1] that we have used for the validation purpose. SAVI is a Canadian distributed test bed for future Internet applications.

To each architecture component, a medium-size OpenStack flavor was allocated with 4 GB of RAM, 2 vCPUs and 40 GB of Disk. For all scenarios, the *CDN component provider* components, i.e. *CDN component deployment manager*, the *microservice deployer*, and the *microservice orchestrator* are deployed in the Toronto site and the components related to the CDN provider domain, i.e. *CDN deployment manager, CDN controller,* and PoDs are deployed on the Vancouver site. We assume the PoDs are equipped with Ubuntu operation system and have Docker platform installed.

For Microservice Orchestrator, Alfresco Activiti[2] is used as a Business Process Management (BPM) solution, which enables the definition of workflows and orchestration plans in Business Process Model Notation (BPMN). It also exposes REST API to external entities to execute the workflows.

All other architecture components are modeled as Restful web services, developed using Java-based Restlet API. The interaction between components is enabled via the Wide Area Network (WAN) - the Internet.

*C. Design of Microservices*

We assume the requested CDN component is a surrogate server with ABR streaming capabilities. This CDN component is decomposed into two microservices: i) a cache node, and ii) an ABR streaming server. Both microservices are modeled as RESTful web services. They expose two REST-based interfaces, one for interacting with control plane components, e.g. microservice orchestrator, or CDN controller and one for data (e.g. content) exchange for customization. Both microservices are implemented using JAVA-based Restlet [3] API. The ABR functionality is implemented in accordance to MPEG-DASH [16] standard. For this functionality, we have used GPAC MP4Box and X264 video codec tools. Both microservice VNFs are packaged in Docker containers and are pushed to DockerHub[4] repository. It should be noted that the cache node is designed in a way that it automatically downloads a sample video content from Dropbox (as a media server) right after it is deployed. A potential design alternative for microservices is the use of pure HTTP as the pillar of the implementation, similar to the modeling done in a previous work [17].

*D. Measurements*

To obtain some insights about the gains associated with the on-the-fly provisioning of CDN components, three sets of experimentations have been done. The first experiment includes measuring the required time for deployment of CDN surrogate server microservices. The second experiment measures the elapsed time for orchestration of the microservices. The third experiment consists of measuring the latency for the process of CDN components provisioning. Table II shows the obtained latency values for CDN component deployment, orchestration and full provisioning process. The measurements are taken for 10 times, and the average and standard deviation is provided.

---

[1] savinetwork.ca
[2] alfresco.com
[3] restlet.com
[4] hub.docker.com

TABLE II. MICROSERVICE DEPLOYMENT, ORCHESTRATION AND FULL-PROVIUSIONING LATENCY

| Experiment | Latency Average (S) | Standard Deviation (S) |
|---|---|---|
| Deployment delay | 6.03 | 0.22 |
| Orchestration delay | 7.97 | 1.08 |
| Provisioning delay | 19.85 | 1.18 |

The deployment latency is measured from the time the deployment request is initiated to the time the acknowledgment for deployment is arrived (Fig. 4, action 5 to 9). The deployment latency is about 6.03 seconds, which highlights the gain of using NFV technology and furthermore, the use of container-based methods for microservice deployment.

The orchestration latency is measured from the time the orchestration request is initiated to the time the acknowledgment for orchestration is arrived (Fig. 4, action10 to 14). The orchestration latency is about 7.97 seconds. The orchestration plan has a significant impact on the orchestration latency. For example in this experiment, the orchestration plan's activities include reading the access information of newly deployed microservices and then contacting each to share the access information of other microservices. This measurement highlights the overhead of using the microservice architectural style which leads to the need for an orchestration mechanism.

The full provisioning process latency is measured from the moment that *CDN deployment manager* sends a request for surrogate server provisioning (Fig. 4, action 4), to the time that newly added surrogate server is integrated into the existing CDN (Fig. 5, action 2). The average of 10 consequent execution is about 19.85 seconds which highlights the gain of on-the-fly provisioning of CDN components compared to the traditional mode of deployments.

V. CONCLUSION & LESSONS LEARNED

This paper proposes an architecture for on-the-fly provisioning of CDN components using NFV and microservice architecture principles. The CDN components are designed in form of microservices and implemented, packaged and deployed using NFV technology. Then the required process is proposed to orchestrate them and integrate them in existing CDNs. A prototype is implemented. Measurements have also been made. The end-to-end CDN component provisioning delay highlights the gain of on-the-fly provisioning of CDN components compared to static deployments. Yet a lesson learned is the overhead induced by the orchestration of newly deployed CDN microservices. The use of choreography instead of orchestration is planned as a future work to examine the orchestration delay. Other potential future works include but not limited to: the design of the proper interfaces between ISP domain and CDN provider domain for PoD information negotiations, and proposing the required algorithms for optimal PoD selections in the pre-deployment phase of provisioning.

ACKNOWLEDGEMENT

This work was supported in part by Ericsson Canada, and the Natural Science and Engineering Council of Canada (NSERC) through the SAVI Research Network.